\documentstyle[12pt,epsfig]{article}

\def\beq{\begin{equation}}
\def\eeq#1{\label{#1}\end{equation}}
\def\eeqn{\end{equation}}

\def\beqa{\begin{eqnarray}}
\def\eeqa#1{\label{#1}\end{eqnarray}}
\def\eeqan{\end{eqnarray}}

\let\bar=\overbar

\def\Dslash{\not{\hbox{\kern-4pt $D$}}}
\def\dslash{\not{\hbox{\kern-2pt $\del$}}}

\def\msb{{\bar{\ssstyle M \kern -1pt S}}}

\def\Title#1{\begin{center} {\Large #1 } \end{center}}

\begin{document}

\begin{flushright}
TUHEP-99-04\\
PDK-741\\
\end{flushright}
\vspace*{0.6cm}

\Title{Atmospheric Neutrinos and the Oscillations Bonanza
\protect\footnote{Plenary talk at the XIX International Symposium on Lepton and 
Photon Interactions at High Energies, Stanford University, August 9-14, 
1999.}}

\bigskip\bigskip


\begin{center}  

{\it W. Anthony Mann\index{Mann, W. Anthony Mann}\\
Department of Physics, Tufts University \\
Medford, MA 02155, USA }
\bigskip\bigskip
\end{center}
 
\begin{quote}
{\footnotesize New observations with atmospheric neutrinos from the 
underground experiments \linebreak SuperKamiokande\index{SuperKamiokande}, 
\ Soudan 2\index{Soudan 2}, \ and MACRO\index{MACRO}, \ together with 
earlier results from Kamiokande\index{Kamiokande}
and IMB\index{IMB}, are reviewed. The most recent 
observations reconfirm aspects of atmospheric flavor content and of 
zenith angle distributions which appear 
anomalous in the context of null oscillations. The anomalous trends, 
exhibited with high statistics in both sub-GeV and multi-GeV data of
the SuperKamiokande water Cherenkov experiment, occur also in event 
samples recorded by the tracking calorimeters. The data are well-described 
by disappearance of $\nu_\mu$ flavor neutrinos arising in oscillations with 
dominant two-state mixing, for which there exists a region in 
($\sin^2 2\theta$, $\Delta m^2$) allowed by all experiments. In a 
new analysis by SuperKamiokande, $\nu_\mu \rightarrow \nu_\tau$ 
\index{$\nu_\mu \rightarrow \nu_\tau$}
is favored 
over $\nu_\mu \rightarrow \nu_{sterile}$ 
\index{$\nu_\mu \rightarrow \nu_s$} as the dominant oscillation
based upon absence of oscillation suppression from matter effects at high 
energies. The possibility for sub-dominant $\nu_\mu \rightarrow \nu_e$ 
\index{$\nu_\mu \rightarrow \nu_e$}
oscillations in atmospheric neutrinos which arises with three-flavor mixing, 
is reviewed, and intriguing possibilities for amplification of this oscillation
by terrestrial matter-induced resonances are discussed. Developments and 
future measurements which will enhance our knowledge of the atmospheric 
neutrino fluxes are briefly noted.}
\end{quote}

\section{Atmospheric Neutrino Beamline}

      We are lucky, you and I, to be born here on planet Earth 
and to have as our birthright the unrestricted use of a splendid neutrino 
beamline. Truly remarkable is that the originating hadronic primary beam, 
namely the cosmic ray flux\index{cosmic ray flux}
of protons and assorted stable nuclei, is 
isotropic to high degree. Moreover the beamline target region, which is the 
terrestrial atmosphere, is very nearly spherically symmetric. Together 
these two attributes ensure that neutrino fluxes of this beamline, in the 
absence of neutrino oscillations, must be up/down symmetric with respect to 
the horizon \cite{Ayres,Fisher}. Consequently, observation of a sizable 
neutrino flux up/down
asymmetry by the user community is clear evidence that new physics is 
happening with neutrinos. The null oscillation up/down symmetry for 
neutrino fluxes is however not complete. There are geomagnetic effects
\index{geomagnetic effects} 
which produce mild distortions in the fluxes of low energy neutrinos and of 
horizontal neutrinos. These distortions, which are latitude-dependent, 
provide useful tests for data verity. The beamline delivers neutrino fluxes 
which are wide-band in $E_\nu$ and which contain the flavors 
$(\nu_\mu + \bar{\nu}_\mu)$ and $(\nu_e + \bar{\nu}_e)$. In the absence of 
oscillations these flavors must occur in the ratio 2:1 for sub-GeV 
neutrinos; for multi-GeV neutrinos the ratio should increase 
gradually with $E_\nu$. At or below the Earth's surface the atmospheric flux 
is about $10^3$ $\nu$'s incident per human body per second \cite{Fisher}, 
an amount which is adequate for 
experimentation but does not pose a radiation safety hazard. The 
neutrino path lengths $L$ which are possible in this beamline range from 20 
km for $\nu$'s incident from the local zenith, to 13,000 km for $\nu$'s 
arriving from the opposite side of the globe. Within the beamline there are 
regions of different, roughly uniform, matter densities. These include the 
Earth's mantle \index{Earth mantle}
(density $\simeq$ 4.5 $g/cm^3$) and the Earth's core \index{Earth core}
(density $\simeq$ 11.5 $g/cm^3$). This arrangement may eventually permit 
experimental strategies to be tried which are akin to utilization of 
regeneration plates in $K^0$ beams. The neutrinos from this ever-running 
beamline give rise to useful reaction final states both in and below any 
detector deployed underground. By investigating the full 
panoply of event types possible with charged current (CC) or neutral 
current (NC) interactions, experimentalists can explore the physics of 
atmospheric neutrinos for incident $E_\nu$ ranging from 100-200 MeV up to 
and exceeding 1000 GeV. 

\section{Oscillation Phenomenology}

      We believe there to be three active neutrinos; there may be sterile 
ones $\nu_s$ as well. For the active neutrinos, the weak flavor
eigenstates $\nu_e$, $\nu_\mu$, and $\nu_\tau$ are related to the mass
eigenstates according to a product involving the unitary mixing matrix 
\index{mixing matrix} $U$:

\begin{equation}
        \left[
                \begin{array}{c}
                \nu_e \\ \nu_\mu \\ \nu_\tau
                \end{array}
        \right]
        =
        \left[
                \begin{array}{ccc}
                U_{e1} & U_{e2} & U_{e3} \\
                U_{\mu1} & U_{\mu2} & U_{\mu3} \\
		U_{\tau1} & U_{\tau2} & U_{\tau3} \\
                \end{array}
        \right]
        \left[
                \begin{array}{c}
                \nu_1 \\ \nu_2 \\ \nu_3
                \end{array}
        \right].
\end{equation}

\noindent
The oscillation probabilities which follow from this can, in principle, 
involve numerous competing processes: 

\begin{equation}
P(\nu_\alpha \rightarrow \nu_\beta) = \delta_{\alpha\beta} -
4 \sum_{i=1}^{3} \sum_{j=i+1}^{3} U_{\alpha i} \cdot U_{\beta i} \cdot
U^{*}_{\alpha j} \cdot U^{*}_{\beta j} \cdot 
\sin^2 \left[ \frac{1.27 \mbox{ } \Delta m^{2}_{ij} \cdot L}{E_\nu} \right].
\end{equation}

\noindent
Fortunately there are cases wherein oscillations decouple
so that the situation is well-described by two-neutrino oscillations, 
for which the mixing matrix \index{mixing matrix} is much simpler:

\begin{equation}
	\left[
		\begin{array}{c}
		\nu_\alpha \\ \nu_\beta
		\end{array}
	\right]
	=
	\left[
		\begin{array}{cc}
		\cos\theta & \sin\theta \\
		-\sin\theta & \cos\theta
		\end{array}
	\right]
	\left[
                \begin{array}{c}
                \nu_1 \\ \nu_2
                \end{array}	
	\right].
\end{equation}

\noindent
The probability for oscillation between the two participating flavors can
then be written using the well-known expression

\begin{equation}
P(\nu_\alpha \rightarrow \nu_\beta) = \sin^2 (2\theta) \cdot
\sin^2 \left[ \frac{1.27 \mbox{ } \Delta m^{2} [\mbox{eV}^2] \cdot L 
\mbox{[km] }}{E_\nu \mbox{[GeV] }} \right].
\end{equation}

\noindent
It is convenient to define the vacuum oscillation length $L_0$:
\index{vacuum oscillation length}

\begin{equation}
L_0 \mbox{[km] } = \pi \left[ \frac{1.27 \mbox{ } \Delta m^{2}}{E_\nu} \right]^{-1}
= 2.47 \frac{E_\nu \mbox{[GeV] }}{\Delta m^2 [\mbox{eV}^2]}.
\end{equation}
\noindent
The oscillation phase can then be expressed as ($\pi L/L_0$).

\section{The Underground Detectors}

Currently there are three underground experiments which are accumulating 
atmospheric neutrino data. The premier detector in this field is 
SuperKamiokande (Super-K)\index{SuperKamiokande}. 
It is a 50 kiloton water Cherenkov detector 
deployed in a configuration of two concentric cylindrical 
volumes. The inner volume is the 22.5 kiloton fiducial region, while 
the surrounding outer volume is used to veto entering tracks and to tag 
exiting tracks. Flavor tagging of events is based upon the  relative 
sharpness or diffuseness of Cherenkov rings, with muon tracks yielding 
sharp rings, and electrons yielding diffuse ones \cite{Kajita}.
The analyzed exposure for Super-K in-detector neutrino reactions reported 
here is from 848 livedays; this corresponds to a whopping 52 fiducial 
kiloton years! 

MACRO \index{MACRO} and Soudan 2 \index{Soudan 2}
are tracking calorimeter detectors. MACRO is a
large-area, planar tracker. It is optimized for tracking in vertical
directions and is sufficiently massive (about 5.3 kilotons) to be effective
as a neutrino detector. Charged particle tracking is carried out using
horizontal layers of streamer tubes with wire and stereo strip readout.
Three horizontal planes and also vertical walls of liquid scintillator
counters provide timing information with resolution of about 0.5 nsec
\cite{Ahlen2}. MACRO has the largest rock overburden of the three
underground experiments, consequently the flux of downgoing muons which can
give rise to backgrounds is lowest at its site \cite{Ambrosio}.

Soudan 2 \index{Soudan 2}
is a fine-grained iron tracking calorimeter of total mass 
963 tons which images non-relativistic as well as relativistic charged 
particles produced in neutrino reactions. The detector operates as a 
slow-drift time projection chamber. Its tracking elements are meter-long plastic 
drift tubes which are placed into the corrugations of steel sheets. The 
sheets are stacked to form a tracking lattice of honeycomb geometry. A 
stack is packaged as a calorimeter module and the detector is 
assembled building-block fashion using these modules. The calorimeter is 
surrounded on all sides by a cavern-liner active shield array of 
two or three layers of proportional tubes \cite{Allison}. The contained 
event sample reported here is obtained from a 4.6 fiducial kton-year 
exposure.

\section{Atmospheric Neutrino Flavor Ratio}

A hypothesis test of long-standing for the existence of anomalous 
behavior of atmospheric neutrinos is ``the flavor ratio'' 
\index{flavor ratio} for which
updated measurements have become available this summer.
Atmospheric neutrinos are produced almost entirely in
pion - muon decay chains initiated by cosmic ray interactions in the
upper atmosphere.  As a consequence, $(\nu_\mu + \bar{\nu}_{\mu})$ versus
$(\nu_e + \bar{\nu}_e)$ neutrino flavor rates occur in a ratio 2:1. 
The underground experiments examine the ratio-of-ratios $R$, which is
$(\nu_\mu + \bar{\nu}_{\mu})/(\nu_e + \bar{\nu}_e)$ from the data, divided by
the same ratio from a Monte Carlo. In the absence of new physics the 
ratio-of-ratios should be unity; and so the degree to which deviation 
from unity is observed is a measure of anomalous behavior of the fluxes.
In actual practice, the experiments measure a related quantity, 
$R^{\prime}$, the ratio of observed event counts. For the Super-K
\index{SuperKamiokande}
water Cherenkov experiment, $R^{\prime}$ is the ratio of single-ring $\mu$-like
to $e$-like events in the data divided by $\mu$-like to $e$-like from the
Monte Carlo \cite{Fukuda6}.  For Soudan 2, \index{Soudan 2}
$R^{\prime}$ is the 
ratio of single-track to single-shower events for the data, divided by the
same ratio from the Monte Carlo \cite{Allison2}.

Here, then, are the latest results from Super-K, \index{SuperKamiokande}
updated to include the 848 day exposure:
For the ``sub-GeV'' sample (with event visible energy $E_{vis}$ $<$ 1.33 GeV),

\begin{displaymath}
           R^{\prime}(\mu\mbox{-ring}/e\mbox{-ring}) = 0.68 \pm 0.02 
\mbox{ (stat.) } \pm 0.05 \mbox{ (syst.) } .
\end{displaymath}

\noindent
For the ``multi-GeV'' sample ($E_{vis} >$ 1.33 GeV),

\begin{displaymath}
            R^{\prime}(\mu\mbox{-ring}/e\mbox{-ring}) = 0.68 \pm 0.04 \pm 0.08 .
\end{displaymath}

From the Soudan 2 \index{Soudan 2}
iron calorimeter there is an updated measurement
based upon contained track and shower events of a 4.6 fiducial kiloton year 
exposure. The events occur mostly within the sub-GeV regime as defined by 
Super-K:

\begin{displaymath}
            R^{\prime}(trk/shwr) = 0.68 \pm 0.11 \pm 0.06.
\end{displaymath}

\begin{figure}[htb]
\centerline{\epsfig{file=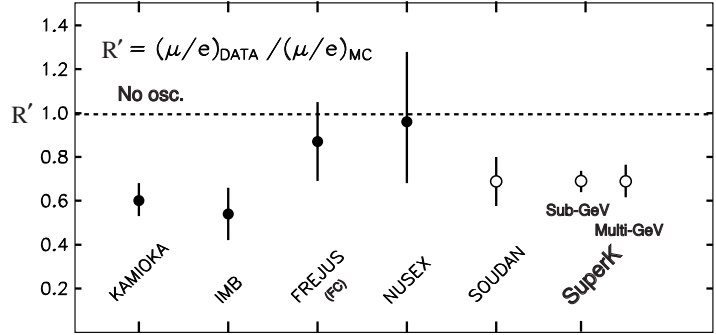,height=1.75in}}
\caption{{\footnotesize Measurements of the atmospheric neutrino flavor
ratio of ratios \protect\cite{Harrison}.}}
\label{fig:fig01}
\end{figure}

      Measurements of the atmospheric flavor ratio \index{flavor ratio}
have been 
accumulating from the underground experiments for more than a decade
\cite{Hirata,Casper,Aglietta2,Berger}. These most recent results 
reconfirm the atmospheric anomaly as first reported years ago by the
water Cherenkov experiments Kamiokande \index{Kamiokande}
and IMB\@. Fig.~\ref{fig:fig01} \index{IMB}
shows that the various $R'$ measurements, by different experiments with 
different techniques and systematics, give a consistent picture. The flavor 
content of the atmospheric neutrino flux is anomalous but in a way that is 
readily understandable, if indeed muon neutrinos are being depleted by 
$\nu_\mu \rightarrow \nu_x$ oscillations over pathlengths which occur in the 
terrestrial beamline.

\section {Zenith Angle Distortions and Super-K Data}

\begin{figure}[htb]
\centerline{\epsfig{file=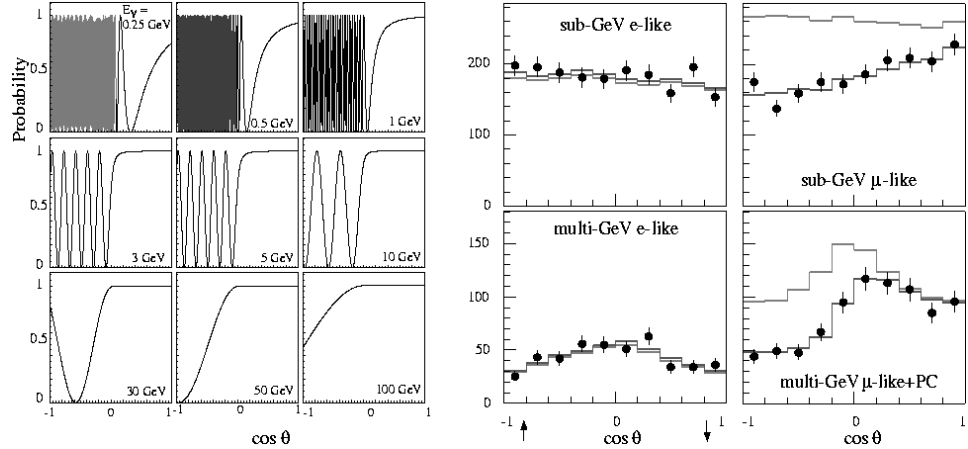,width=5.0in}}
\caption{{\footnotesize {\bf a (left)}: Survival probability curves for
mono-energetic, isotropic fluxes
of muon neutrinos with $\nu_\mu \rightarrow \nu_\tau$ and $\sin^2 2\theta = 1.0$,
$\Delta m^2 = 5 \times 10^{-3}$ eV$^2$ for nine values of $E_\nu$. 
{\bf b (right)}:
Super-K distributions for $\cos \theta_z$ single ring $e$-like and
$\mu$-like events. Expectations for null oscillation and for 
$\nu_\mu \rightarrow \nu_\tau$ are shown by gray-line and solid-line 
histograms respectively.}}
\label{fig:fig02}
\end{figure}

      To elicit the pathlength $L$ dependence which, in an oscillation 
scenario, will correlate with ($\nu_\mu + \bar{\nu}_\mu$) disappearance, we 
consider the distributions of neutrino zenith angle which have
been obtained for fully contained (FC) and for partially contained (PC)
events in the SuperKamiokande \index{SuperKamiokande} experiment. 
In evaluating zenith angle distributions \index{zenith angle distributions}
and also flavor ratios, it is useful to keep in mind
trends which are shown by the survival probability curves in 
Fig.~\ref{fig:fig02}a for $\nu_\mu$ neutrinos \cite{Stenger}.
The curves depict the probability for $\nu_\mu \rightarrow \nu_\mu$ 
from an atmospheric flux for which $\cos \theta_z$ at 1.0 is vertically 
downgoing and $\cos \theta_z$ at -1.0 is vertically upgoing.  The curves 
are drawn for ``representative'' $\nu_\mu \rightarrow \nu_x$
parameter settings which we use again in paragraphs below, namely
$\sin^2 2\theta = 1.0$ and $\Delta m^2 = 5\times 10^{-3}$ eV$^2$.  
The oscillation pattern in Fig.~\ref{fig:fig02}a evolves in a regular 
way with increasing energy of the neutrino. For $E_\nu$ of 250 MeV, the 
first oscillation swing severely depletes the downward-going flux, and 
rapid oscillations deplete the flux incident from below-horizon; 
the net result is a substantial average depletion at all incident 
angles.  At energies above 1 GeV however the depletion moves almost
entirely to the $\nu_\mu$ flux incident from below-horizon, and this situation 
remains for $E_\nu$ increasing to 30 GeV\@. At higher $E_\nu$ the pattern 
shifts to beyond range, and $\nu_\mu$ depletion ceases because our planet 
is not big enough to accomodate the first oscillation swing.

	Distributions showing ten bins in $\cos \theta_z$ for events of the 
848-day Super-K \index{SuperKamiokande}
exposure are given in Fig.~\ref{fig:fig02}b. The 
$\nu_e$ and $\nu_\mu$ flavor samples are shown separately and are 
subdivided according to $E_{vis}$. The $\nu_e$ events show no angular 
distortion in either the sub-GeV or multi-GeV regimes.  In striking 
contrast the $\nu_\mu$ samples show large regions of disappearance, 
the samples being depleted relative to expectations of the null oscillation 
Monte Carlo (gray-line histograms). The depletions exhibit dependence on 
zenith angle and therefore on path length $L$. Additionally, the depleted
regions are of different character in the sub-GeV and multi-GeV sets.
At sub-GeV energies the $\mu$-like events appear depleted at all angles 
including those with incidence from above horizon. At multi-GeV energies 
however, the depletion is mostly restricted to incidence from 
below-horizon. Although the correlation between the final state 
lepton and the initial neutrino direction is relatively poor for sub-GeV 
compared to multi-GeV data, nevertheless the trend is suggestive of a 
dependence on $E_\nu$ for $\nu_\mu$ flavor disappearance. As shown by the 
solid-line histograms superposed in Fig.~\ref{fig:fig02}b, the zenith 
angle distortions of the $\nu_\mu$ flavor samples are well-described by a 
fit of two-state $\nu_\mu \rightarrow \nu_\tau$ 
\index{$\nu_\mu \rightarrow \nu_\tau$} neutrino oscillations 
(discussed below). Contrastingly, the $\nu_e$ samples are in agreement with 
the null oscillation Monte Carlo (MC) to a degree which is perhaps 
disappointing. With multi-GeV $\nu_e$'s which presumably traversed the 
Earth's core, for example, no irregularity is apparent; there are no hints 
anywhere to suggest $\nu_\mu \rightarrow \nu_e$ 
\index{$\nu_\mu \rightarrow \nu_e$}
oscillations.

\begin{figure}[htb]
\centerline{\epsfig{file=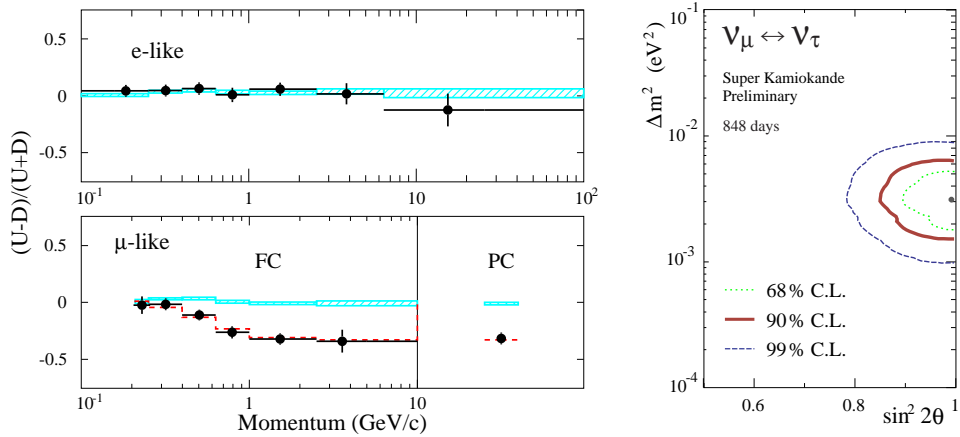,width=5.0in}}
\caption{{\footnotesize {\bf a (left)}: Zenith angle asymmetry versus event momentum for single ring $e$-like
and $\mu$-like events of Super-K. {\bf b (right)}:
Super-K allowed region in the neutrino oscillation parameter space
for $\nu_\mu \rightarrow \nu_\tau$ mixing, based upon the $\chi^2$ fit
to FC and PC single ring event distributions.}}
\label{fig:fig03}
\end{figure}

      The depletion of $\nu_\mu$ neutrinos can be shown in an informative
way by plotting the asymmetry in zenith angle 
\index{zenith angle asymmetry} as a function of
event momentum as in Fig.~\ref{fig:fig03}a. The asymmetry $A$ is defined
$A = (U-D)/(U+D)$, where $U$ is the number of events with upward incidence at
angles $\cos \theta > 0.2$ and $D$ equals events with downward incidence 
at angles $\cos \theta < -0.2$.  For the single-ring $e$-like events, 
$A$ $\simeq$ 0 at all momenta.  For the $\mu$-like events however, $A$ becomes
increasingly negative, there being a dearth of upward-going versus 
downward-going neutrinos which becomes more pronounced with increasing 
momentum. For the multi-GeV sample the value of $A$ is 
$-0.32 \pm 0.04 \pm 0.01$, which is nearly eight standard deviations from 
zero asymmetry.

      For the fully-contained and partially contained single-ring data just
shown, the Super-K \index{SuperKamiokande}
collaboration uses a $\chi^2$ function to determine the 
oscillation parameters of two-state mixing:

\begin{equation}
\chi^2 (\sin^2 2\theta, \Delta m^2, \vec{\varepsilon}) =
\sum_{i=1, 70} \frac {(N^i_{data} - N^i_{MC})^2}{\sigma^2_i} +
\sum_j \frac{\varepsilon^2_j}{\sigma^2_j}.
\end{equation}

\noindent
For this purpose the $\mu$-like and the $e$-like samples are 
sub-divided using five bins in $\cos \theta_z$ and seven bins in
momentum. The $\chi^2$ is the sum of data minus MC expectation squared 
over the 70 bins, where the MC is a function of the oscillation parameters 
$\sin^2 2\theta$, $\Delta m^2$, and parameters $\varepsilon_j$ which allow 
for systematic effects. The $\varepsilon_j$ include the parameter $\alpha$ 
which appears in the flux normalization factor (1 + $\alpha$).
At each point in the plane of $\sin^2 2\theta$ and $\Delta m^2$, the $\chi^2$
is minimized with respect to the $\varepsilon_j$ parameters;
the minimum $\chi^2$ point (best fit) is then obtained.  Contours for
allowed regions \index{allowed regions}
at 68\%, 90\%, and 99\% CL are obtained on the basis of
$\chi^2 - \chi^2_{minimum}$ as shown in Fig.~\ref{fig:fig03}b.
The $\nu_\mu \rightarrow \nu_\tau$ 
\index{$\nu_\mu \rightarrow \nu_\tau$} oscillation best fit yields $\chi^2$ =
55/67 d.o.f. and fares much better than the null oscillation fit having 
177/69 d.o.f. At the best fit point the oscillation parameter values 
are $\sin^2 2\theta = 0.99$ and $\Delta m^2 = 3.1 \times 10^{-3}$ [eV$^2$];
the MC flux normalization is shifted upwards ($\alpha$ = +0.05) relative to 
the absolute rate based upon the Honda {\it et al}. fluxes for the Super-K 
\index{SuperKamiokande} site
\cite{Honda2}. It is comforting to see that $\sin^2 2\theta$ from the best fit 
with Super-K FC and PC events now occurs in the physical region, for this has not 
always been the case in the past.

\begin{figure}[htb]
\centerline{\epsfig{file=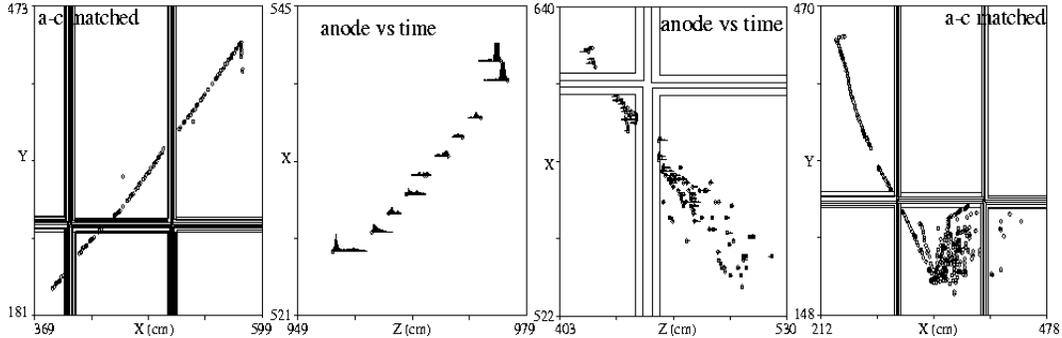,width=5.5in}}
\caption{{\footnotesize Two track-plus-recoil events, a single shower event, 
and a $\bar{\nu}_\mu$ flavor multiprong event, recorded in Soudan 2.}}
\label{fig:fig04}
\end{figure}

\section{Contained Events in Soudan 2}

	Concerning evidence for neutrino oscillations carried by 
in-detector neutrino interactions, a ``second look'' is afforded by 
the fully contained track, shower, and multiprong events recorded 
by Soudan 2. \index{Soudan 2} 
Projected images of ``typical'' data events are shown in 
Fig.~\ref{fig:fig04}; these include two examples of muon tracks with 
companion recoil protons ($\nu_\mu$ quasi-elastics), a shower event
($\nu_e$ quasi-elastic), and a $\bar{\nu}_\mu$-flavor multiprong.

      The approach taken by Soudan 2 is to isolate a sub-sample of events 
for which $L/E_\nu$ \index{$L/E_\nu$}
can be measured with good resolution on an event-by-event 
basis, thereby allowing the oscillation analysis to be carried out directly 
using $L/E_\nu$ distributions.  In a fine-grain tracking calorimeter, 
$E_\nu$ can be reliably 
estimated based upon $E_{vis}$. To ensure good resolution for ascertaining 
the incident neutrino direction, quasi-elastic single track and shower 
events are selected which have measurable recoil protons. Otherwise, in the 
absence of a visible recoil, the lepton energy is required to exceed 600 
MeV. Multiprong events are also selected, provided that 
$E_{vis} >$ 700 MeV and $|\Sigma \vec{p}_{vis}| >$ 450 MeV/$c$ and
$P_{lept} >$ 250 MeV/$c$. The momentum requirements improve the 
resolution of neutrino direction and ensure reliable flavor-tagging for 
charged current events (success probability $>$ 0.92). 
For the $\nu_\mu$ ($\nu_e$) sample, $\Delta E/E$ is 20\% (23\%).
For pointing of the event along the
original neutrino direction, the resolutions are of order 20-30 degrees
which is quite respectable for a sub-GeV data set \cite{Gallagher}.

\begin{figure}[htb]
\centerline{\epsfig{file=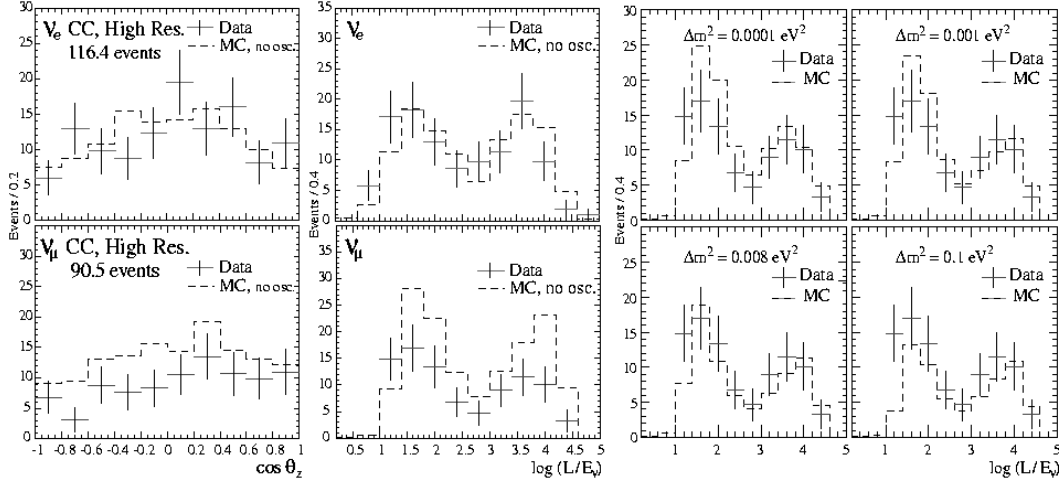,width=5.5in}}
\caption{{\footnotesize {\bf a (left)}: Distributions in $\cos \theta_z$ for 
$\nu_e$ and $\nu_\mu$ flavor
event samples. Data (crosses) are compared to the null oscillation MC
(dashed histogram) where the MC has been rate normalized to $\nu_e$ data. 
{\bf b (middle)}: Distributions of $\log (L/E_{\nu})$ for $\nu_e$ and $\nu_\mu$
charged current events compared to the neutrino MC with no oscillations;
the MC has been rate-normalized to the $\nu_e$ data.
{\bf c (right)}: Comparison of $L/E_{\nu}$ distribution for $\nu_\mu$ data
(crosses)
and expectations from neutrino oscillations for four $\Delta m^2$ values,
with $\sin^2 2\theta = 1.0$.}}
\label{fig:fig05}
\end{figure}

Zenith angle distributions for the resolution-enhanced (HiRes)
$\nu_e$ and $\nu_\mu$ samples are 
shown in Fig.~\ref{fig:fig05}a, where the MC rates have been normalized 
to the observed number of $\nu_e$ events. For the $\nu_e$ 
sample, the zenith angle distribution follows the shape predicted by the MC 
without oscillations. The corresponding $\nu_\mu$ data distribution however 
consistently falls below the MC expectation. The dearth is mild but 
discernible for $\nu$ incidence above the horizon and becomes more 
pronounced with below-horizon incidence. These features are in agreement with 
those exhibited with much higher statistical weight by the sub-GeV FC 
single-ring events of Super-K. \index{SuperKamiokande}
 
Distributions in $\log L/E_\nu$ for the HiRes $\nu_e$ and $\nu_\mu$ 
samples are shown in Fig.~\ref{fig:fig05}b wherein the data (crosses) are 
compared to the null oscillation MC. 
The peaks at low $\log L/E_\nu$ are populated by down-going
neutrinos incident from above-horizon; the lower flux central regions are 
populated by horizontal neutrinos, while the peaks at higher $\log L/E_\nu$
contain neutrinos traveling upward through the Earth.
To within statistical fluctuations, the $\nu_e$ sample
follows the null oscillation MC expectation. For the $\nu_\mu$ sample,
there is a depletion which pervades the entire
up-going region and extends into the down-going flux,
subsiding only in the lowest $L/E_\nu$ bins which contain the most
vertically down-going events.

\begin{figure}[htb]
\centerline{\epsfig{file=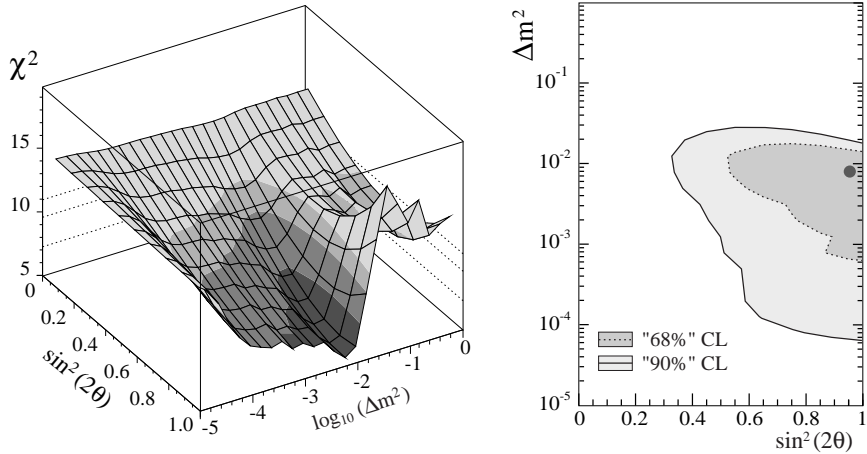,width=4.5in}}
\caption{{\footnotesize {\bf a (left)}: $\chi^2$ ``map'' over the $\Delta m^2$, $\sin^2 2\theta$ plane;
the MC normalization is also allowed to adjust. Contour boundaries are
approximately 68\% CL, 90\% CL, and 95\% CL based upon $\chi^2_{min}$ plus
2.30, 4.61, and 5.99 respectively. 
{\bf b (right)}: 
Soudan 2 allowed regions for $\nu_\mu \rightarrow \nu_x$
oscillations at approximately 68\% and 90\% CL.}}
\label{fig:fig06}
\end{figure}

The $\log L/E_\nu$ distributions from data can be fitted to 
oscillation-weighted MC events using a $\chi^2$ constructed similarly to 
the function utilized by Super-K. \index{SuperKamiokande}
An exploratory matchup is shown in 
Fig.~\ref{fig:fig05}c, for which $\sin^2 2\theta$ is set to 1.0 and the
$\nu_\mu$ data is displayed together with weighted MC distributions for
$\nu_\mu \rightarrow \nu_x$ oscillations at four different $\Delta m^2$ 
values. At $\Delta m^2 = 10^{-4}$ eV$^2$ the oscillation solution 
exceeds the data in the down-going hemisphere. At 
$\Delta m^2 = 10^{-3}$ eV$^2$ the matchup improves, and at
$\Delta m^2 = 8\times10^{-3}$ eV$^2$ the oscillation solution 
follows the data rather well. However 
$\Delta m^2 = 10^{-1}$ eV$^2$, is ``too far'' - the oscillation 
solution falls below the data in the down-going hemisphere and in the 
up-going hemisphere as well. This sequence illustrates key 
features of the $\chi^2$ mapping of the ($\Delta m^2$, $\sin^2 2\theta$) 
plane shown in Fig.~\ref{fig:fig06}a. The best fit lies in the darkened 
basin region of the contour. The boundaries of the different shaded areas 
correspond to regions allowed \index{allowed regions}
at approximately 68\%, 90\%, and 95\% CL.

         In the contour map there appears a
`ridge of improbability' at $\Delta m^2$ lying just above
the best fit `basin'.  This region corresponds to oscillation solutions 
for which the first oscillation swing should create a depletion in the 
downward-going $\nu_\mu$ flux.  No such depletion  occurs in the data and  since
the events have sufficient directional resolution to show it if it occurred, 
the $\chi^2$ becomes large there.  The projection of the $\chi^2$ contours onto
the ($\Delta m^2$, $\sin^2 2\theta$) plane is shown in 
Fig.~\ref{fig:fig06}b. 
The minimum $\chi^2$ point is at $\sin^2 2\theta = 0.95$ and
$\Delta m^2 = 0.8 \times 10^{-2}$ eV$^2$.
The flux normalization, which is allowed to vary in the fitting, is reset 
at the minimum $\chi^2$  point to 0.82 times the absolute event rate based 
upon the Monte Carlo. The Monte Carlo utilizes the 1989
Bartol flux calculation for the Soudan \index{Soudan} site \cite{Gaisser}.

\begin{figure}[htb]
\centerline{\epsfig{file=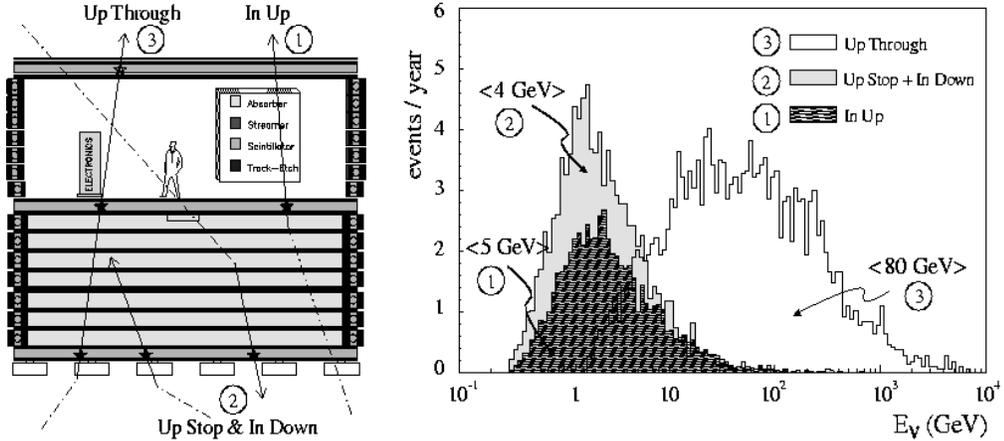,width=5.2in}}
\caption{{\footnotesize {\bf a (left)}: Cross section sketch of the MACRO detector, illustrating the
event topologies of partially contained and through-going $\nu_\mu$ samples.
{\bf b (right)}: 
Distributions of parent neutrino energies for the three neutrino
samples of MACRO.}}
\label{fig:fig07}
\end{figure}

\section{Partially Contained Events in MACRO}

	Although the Soudan results are in general agreement with the 
neutrino oscillation effects reported by Super-K, \index{SuperKamiokande}
they do not at present 
confirm the striking depletion in upgoing muon neutrinos shown by 
Super-K's  multi-GeV events. Fortunately, event samples 
which provide another, independent viewing of the multi-GeV $E_\nu$ regime
are being accumulated in the MACRO \index{MACRO}
experiment \cite{Surdo}.

\begin{figure}[htb]
\centerline{\epsfig{file=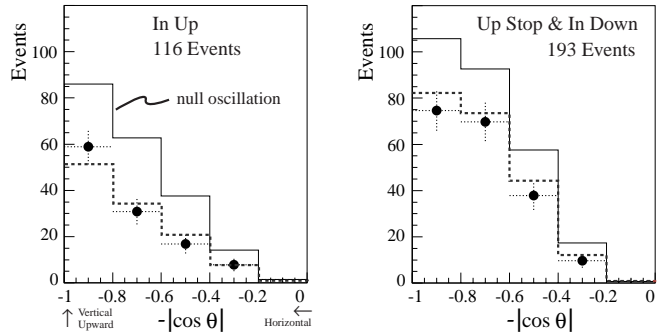,height=1.75in}}
\caption{{\footnotesize Distributions in $\cos \theta_z$ for MACRO partially contained 
events.
The data (solid circles) are seen to fall below the null oscillation
expection in every bin of both samples.}}
\label{fig:fig08}
\end{figure}

      Fig.~\ref{fig:fig07}a shows the three $\nu_\mu$ CC event categories
studied by MACRO. \index{MACRO} These include: {\it i)} ``In-Up'' events, 
which are $(\nu_\mu + \bar{\nu}_\mu)$ reactions occurring inside
the detector creating muons which exit through the top; 
{\it ii)} ``Up-Stop'' and ``In-Down'' events which are classified on
the basis of topology (timing information not available)
and which are analyzed together; 
and {\it iii)} ``Up-Through muons'' which are initiated by high energy 
$(\nu_\mu + \bar{\nu}_\mu)$ interactions below the detector creating muons 
which traverse the detector from bottom to
top. Parent neutrino energy distributions for each of the three event 
categories are shown in Fig.~\ref{fig:fig07}b. 

The In-Up events, and also 
the Up-Stop plus In-Down events, probe the multi-GeV $E_\nu$ regime;
the mean $E_\nu$ values for these two MACRO \index{MACRO}
samples are 5 GeV and 4 GeV respectively.
Zenith angle distributions \index{zenith angle distributions}
for these partially contained samples  
are shown in Fig.~\ref{fig:fig08}. The event 
populations are binned in $\cos \theta_z$ from the horizontal 
at $\cos \theta_z = 0.0$, to vertically upward at zenith cosine -1.0.  For the
116 events of the In-Up sample shown in Fig.~\ref{fig:fig08}a, 
the data fall below the null oscillation Monte Carlo in every bin. (Note that
the acceptance for this planar calorimeter is relatively lower for horizontal
directions.) The ratio of In-Up events observed to the MC prediction, is $0.57 \pm 0.16$.
Thus the In-Up sample exhibits the large-scale depletion for multi-GeV 
upward going events seen in the Super-K data.  The data are seen to 
distribute in accord with the oscillation best fit based upon MACRO 
Up-Through muons which is described in the next Section. A similar trend
is observed with the Up-Stop and In-Down sample shown in 
Fig.~\ref{fig:fig08}b.  Since the latter sample contains 
roughly equal portions of Up-Stop events which are fully 
oscillating and of In-Down events from above horizon which 
are not oscillating, the amount of depletion relative to null 
oscillation is reduced compared to that of the In-Up sample.

\section{Upgoing Muons in Super-K and MACRO}

	There are two event samples which are initiated by  
below-detector $(\nu_\mu + \bar{\nu}_\mu)$
interactions, namely upward stopping muons 
and upward through-going muons\index{upward muons},
and they represent two different portions of the neutrino spectrum.
This can be seen from comparison of $E_\nu$ distributions (2) and (3)
of Fig.~\ref{fig:fig07}b, which roughly characterize the muon samples in 
Super-K \index{SuperKamiokande}
as well as in MACRO. \index{MACRO}
The $E_\nu$ spectrum (2) which produces 
upward stopping muons is distinctly lower,
with the bulk of the spectrum lying below 40 GeV. The 
different $E_\nu$ regimes give rise to rather different 
oscillation behavior, as can be seen by evaluation of the phase angle in Eq. 
(4) at our nominal $\nu_\mu \rightarrow \nu_\tau$ 
\index{$\nu_\mu \rightarrow \nu_\tau$} parameter values 
$\sin^2 2\theta = 1.0$ and $\Delta m^2 = 5\times10^{-3}$ eV$^2$. At $E_\nu$
= 40 GeV, the vacuum oscillation length $L_0$ equals 1.5 Earth diameters. 
Recall that $L_0$ is proportional to $E_\nu$ and that the oscillation phase 
is $\pi L/L_0$. Then for $E_\nu$ much larger than 40 GeV as in the case 
for many through-going muon events, $L_0$ exceeds the Earth's 
diameter. The result is that $\nu$'s initiating through-going muons 
generally have small oscillation phase angles and hence give rise to 
low oscillation probabilities. On the other hand, for
up-stopping muons, $E_\nu <$ 40 GeV and neutrino $L_0$ values are less 
than $L$ values so that sizable phases and large oscillation probabilities 
frequently occur. 

\begin{figure}[htb]
\centerline{\epsfig{file=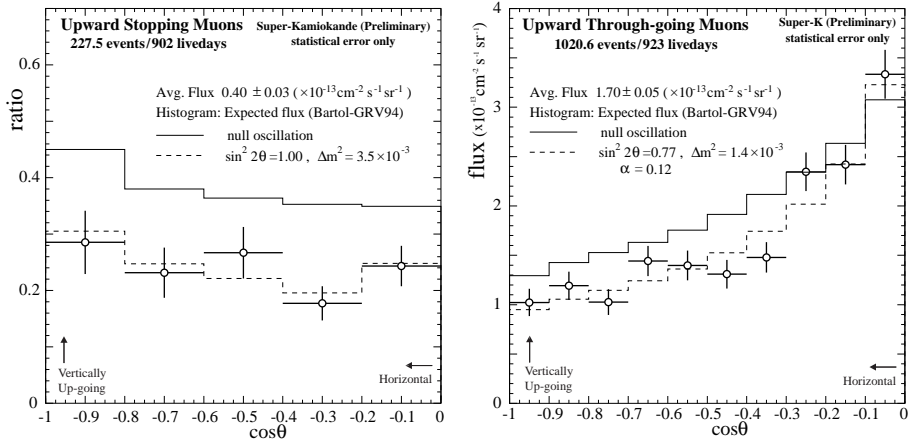,width=4.75in}}
\caption{{\footnotesize {\bf a (left)}: Ratio of upward-stopping to through-going muons versus 
$\cos \theta_z$ observed in Super-K. The data ratio falls below the null 
oscillation expectation for each of the five below-horizon bins. 
{\bf b (right)}: 
The upward through-going muon flux versus zenith angle observed in
Super-K. The shape of the data distribution (open circles) differs
from the null oscillation expectation (solid histogram) and is better
described by two-state oscillations (dashed histogram).}}
\label{fig:fig09}
\end{figure}

	The expectation that oscillation will occur in relatively different 
proportions in up-stopping versus upward through-going muon samples has been 
examined by Super-K. This is done by measuring the 
upward-stopping to up-through ratio of muon fluxes. In the presence of the 
two-state mixing 
inferred from the in-detector samples of Sect. 6, the muon stop/thru ratio 
should fall below the MC prediction for null oscillation.
Fig.~\ref{fig:fig09}a shows the stop/thru ratio plotted versus
$\cos \theta_z$ for muons incident from the horizontal 
($\cos \theta_z = 0.0$) to those most vertically upgoing
($\cos \theta_z = -1.0$). The observed stop/thru ratio is 0.24
$\pm$ 0.02 which is 2.8 standard deviations below the null oscillation 
expectation of 0.37 $\pm$ 0.05.

\begin{figure}[htb]
\centerline{\epsfig{file=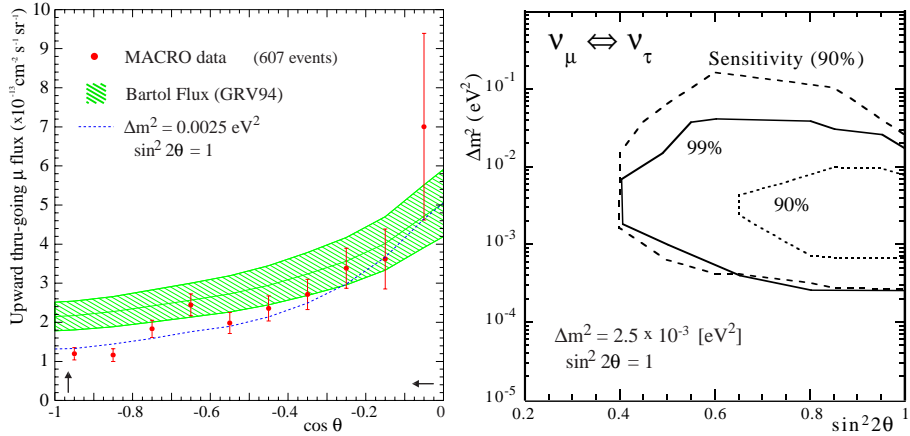,width=4.75in}}
\caption{{\footnotesize {\bf a (left)}: The angular distribution of upward through-going muons observed in
MACRO. The data distribution (solid circles) differs from the null
oscillation expectation (shaded band) in shape and in rate.
{\bf b (right)}: 
The neutrino oscillation allowed regions obtained by MACRO from
the upward through-going muons. Confidence
level and experimental sensitivity boundaries are calculated using the
Feldman-Cousins method.}}
\label{fig:fig10}
\end{figure}

      Additional information can be gleaned from the upward 
through-going muons alone. These events arise from neutrino interactions which 
have the highest range for parent $E_\nu$. To see how 
oscillations affect this sample, consider $E_\nu$ at 100 GeV, which is 
the mean energy estimated for Super-K \index{SuperKamiokande}
events. (However the 
distribution of parent energies is broad and extends above 1000 GeV.)
At our representative parameter settings, the vacuum oscillation length is 
approximately $1.2 \pi$ times the Earth diameter; 
consequently the phase angle of the 
flavor oscillation probability is $L/1.2$ in units of Earth diameter.
For horizontal muons the flight paths of parent neutrinos are of order 
500 km or 0.04 Earth diameters, and so the 
neutrino phase angles will be too small 
to induce significant oscillation probability. However for vertical muons
the neutrino paths $L$ become comparable to the Earth's diameter, and the 
phase angles become sufficiently
large that rapid oscillation swings ensue.  ( For $E_\nu < 100$ GeV, 
oscillations will also occur for muons incident away from the vertical. )

     Available for this conference is an updated through-going 
muon sample exceeding one thousand events from Super-K, 
\index{SuperKamiokande}
the zenith angle
distribution of which is shown in Fig.~\ref{fig:fig09}b. (This sample is 
noticeably larger than the one published this spring from a
537 day exposure \cite{Fukuda}.)
Fig.~\ref{fig:fig09}b exhibits the trends implied by oscillations for this 
sample:  For bins which are just below the horizon, the 
data agree with the Monte Carlo expectation for null oscillation.
For the bins below $\cos \theta_z = -0.3 \mbox{ }$ however, the data fall 
below the null oscillation prediction and this trend continues with muons 
of more vertical inclination. That is, the shape of the zenith angle 
distribution of these upward throughgoing muons, and their overall flux rate 
as well, deviate significantly from null oscillation and agree with 
expectations from $\nu_\mu \rightarrow \nu_\tau$ 
\index{$\nu_\mu \rightarrow \nu_\tau$} mixing as inferred from 
in-detector neutrino interactions. 

      The same features are seen in the angular distribution of upward
through-going muons recorded by MACRO \index{MACRO} \cite{Ahlen}
as shown in Fig.~\ref{fig:fig10}a.
The ratio of data to the MC prediction for MACRO is
$0.74 \pm 0.03 \pm 0.04 \pm 0.12$;
the last term containing the largest uncertainty
reflects limited knowledge of  
the absolute neutrino flux and of deep inelastic neutrino cross sections.

	Fig.~\ref{fig:fig10}b shows the allowed region 
\index{allowed regions} of 
the oscillation parameters obtained by MACRO \index{MACRO} based upon 
the upward through-going muon sample.
The allowed region is  calculated using the
Feldman-Cousins method \index{Feldman-Cousins method}
\cite{Feldman}.  For the MACRO 
data, the minimum $\chi^2$ point ($\chi^2$ of 10.6) is in the unphysical
region at $\sin^2 2\theta = 1.5$ .  To clarify the situation,
the experimental sensitivity at 90\% CL is also provided. This is the 90\% CL
contour that would have been obtained had the data coincided with the 
oscillation MC expectation at the nearest point inside the allowed region
($\sin^2 2\theta = 1.0$; $\chi^2 = 12.5$). 

\section{Best Fits for $\sin^2 2\theta$ and $\Delta m^2$}

\begin{figure}[htb]
\centerline{\epsfig{file=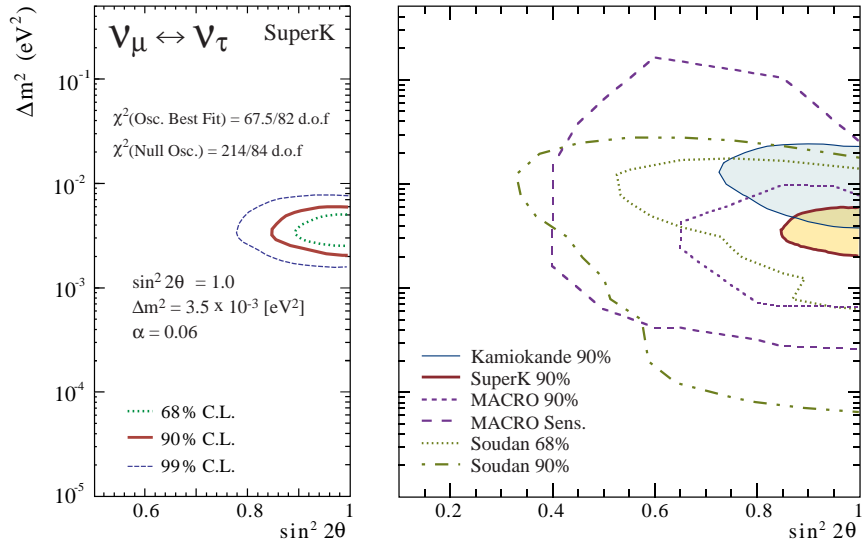,width=4.5in}}
\caption{{\footnotesize {\bf a (left)}: Allowed regions obtained by Super-K based upon $\chi^2$ fitting to 
FC and PC single ring events, plus upward stopping muons, plus upward 
through-going muons. 
{\bf b (right)}: 
Oscillation parameter allowed
regions from Kamiokande (thin-line boundary), Super-K (thick-line
boundary), MACRO (dashed boundaries), and Soudan 2 (dotted and dot-dashed
boundaries).}}
\label{fig:fig11}
\end{figure}

Available for this Symposium is a new `all data fit' by Super-K
\index{SuperKamiokande} which 
determines the $\nu_\mu \rightarrow \nu_\tau$ 
\index{$\nu_\mu \rightarrow \nu_\tau$} allowed region using a
$\chi^2$ summed over all neutrino samples
including FC + PC events (848 livedays) plus up-throughgoing muons 
(923 livedays) plus up-stopping muons (902 livedays).
As shown in Fig.~\ref{fig:fig11}a, 
regions in the $\sin^2 2\theta$, $\Delta m^2$ 
plane allowed by this fit are the most restrictive ones ever obtained. 
The $\chi^2$ for the
oscillation best fit is 67.5 for 82 d.o.f., to be compared with 214 for 84
d.o.f. for null oscillation.  The best fit values are
$\sin^2 2\theta = 1.0$ and $\Delta m^2 = 3.5 \times 10^{-3}$ eV$^2$ with
flux normalization parameter $\alpha$ = +0.06.

In order to gauge the overall consistency of atmospheric neutrino 
observations, the various oscillation-parameter allowed regions 
\index{allowed regions} obtained by 
each of the underground experiments have been assembled in 
Fig.~\ref{fig:fig11}b. Along with the 90\% CL region from the Super-K 
\index{SuperKamiokande} all 
data fit discussed above, we include the final 90\% CL region reported by 
Kamiokande \index{Kamiokande} (thin-line boundary) \cite{Hatakeyama}. 
From MACRO, \index{MACRO}
we show the 90\% CL region and include the experimental 
sensitivity at 90\% CL (dashed boundaries). For Soudan 2, \index{Soudan 2}
we plot the region 
allowed at 68\% and 90\% CL. Fig.~\ref{fig:fig11}b
shows that, for two independent water Cherenkov detectors and for 
two quite different tracking calorimeters, there is a region of oscillation 
parameter values which is acceptable to all experiments. Now ``Fools rush 
in where Angels fear to tread'', it is often said. And concerning the 
significance of Fig.~\ref{fig:fig11}b the Angels will urge caution, for 
the atmospheric data show neutrino disappearance only - oscillation 
appearance has yet to be shown.  Nevertheless, this Fool cannot resist 
the rush: I propose to you that congratulations are in order for the 
researchers of Kamiokande and of Super-K and more generally, for the 
non-accelerator underground physics community. For 
Fig.~\ref{fig:fig11}b, Ladies and Gentlemen, is the portrait of a 
Discovery - the discovery of neutrino oscillations with two-state mixing.

\section{Dominant and Sub-Dominant Two-State Mixing}

Assuming that muon neutrinos oscillate into other flavor(s), with nearly
maximal mixing and with $\Delta m^2$ in the range $10^{-2}$ 
to $10^{-3}$ eV$^2$, it is of interest to consider what flavors are involved in the dominant
two-state oscillation, and in other possible sub-dominant oscillations.
That $\nu_\mu \rightarrow \nu_e$ \index{$\nu_\mu \rightarrow \nu_e$}
could be the dominant mode for $\nu_\mu$ 
disappearance is ruled out by the CHOOZ \index{CHOOZ}
reactor experiment. CHOOZ
has established a limit on $\bar{\nu}_e$ disappearance \cite{Apollonio}; 
by CPT symmetry, this limit implies that $\nu_e$ neutrinos
do not disappear, or at least not in a parameter regime which is relevant to the
atmospheric flux.

	Since it is generally believed that $\nu_\mu \rightarrow \nu_\tau$
\index{$\nu_\mu \rightarrow \nu_\tau$} oscillation is 
the dominant mode, it is natural to ask: Where are the $\nu_\tau$ events? 
In the dominant $\nu_\mu \rightarrow \nu_\tau$ scenario, about 0.9 charged 
current $\nu_\tau$ events per kiloton year exposure can be expected to 
occur in an underground detector \cite{Messier}.
Then, in exposures reported at this Symposium, we would expect Super-K 
\index{SuperKamiokande}
to have recorded $\approx 47$ FC or PC charged current $\nu_\tau$ events
and Soudan 2 \index{Soudan 2}
to have recorded about 4 events. These events will be up-going 
but otherwise indistinguishable from energetic NC events, and so there is 
little hope that $\nu_\tau$ reactions can be isolated by the on-going 
atmospheric neutrino experiments. It would be heartening to see a few 
unambiguous tau-neutrino interactions - even from an accelerator experiment! 
On this, hopes are placed with the $\nu_\tau$ candidates recorded by the 
DONUT \index{DONUT}
hybrid emulsion experiment at Fermilab \cite{Kafka}.

	Although dominant $\nu_\mu \rightarrow \nu_\tau$ 
\index{$\nu_\mu \rightarrow \nu_\tau$} is unlikely to be
confirmed anytime soon via $\nu_\tau$ appearance, progress has been made 
by Super-K \index{SuperKamiokande}
towards eliminating the remaining competition which is 
$\nu_\mu \rightarrow \nu_{sterile}$ \index{$\nu_\mu \rightarrow \nu_s$}
oscillations. Now sterile neutrinos, by 
definition, do not interact with normal matter via neutral currents,
a fact which has consequences currently being examined by Super-K. Firstly, 
it follows that sterile neutrinos cannot produce single $\pi^0$ events
since these are NC reactions: $\nu_s N \not\rightarrow \nu_s N \pi^0$.
Then, relative to the $\nu_\mu \rightarrow \nu_\tau$ scenario, 
$\nu_\mu \rightarrow \nu_{s}$
will result in fewer single $\pi^0$ events, and the relative dearth of these 
final states will be in up-going directions \cite{Learned}. Unfortunately, 
cross sections for these NC reactions have large uncertainties, a situation 
which hinders the isolation of a depletion which is demonstrably significant
\cite{Geiser}.

\begin{figure}[htb]
\centerline{\epsfig{file=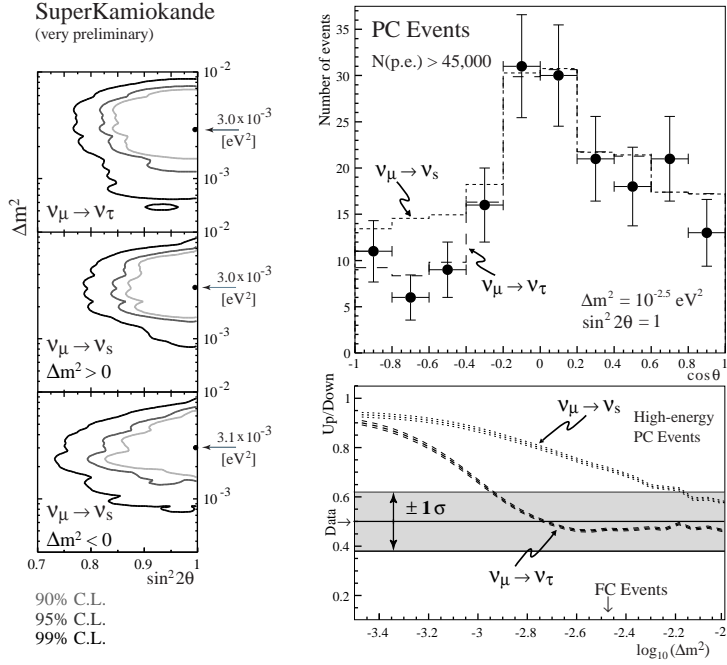,height=3.5in}}
\caption{{\footnotesize {\bf a (left)}: Allowed regions for $\nu_\mu \rightarrow \nu_\tau$ and for
$\nu_\mu \rightarrow \nu_{sterile}$ two-state mixing scenarios, obtained
from FC single ring events of Super-K. 
{\bf b (right,top)}: 
Zenith angle distribution for selected energetic
PC events of Super-K (solid circles); histograms show the
$\nu_\mu \rightarrow \nu_\tau$ and $\nu_\mu \rightarrow \nu_{s}$
solutions of Fig.~\ref{fig:fig12}a.
{\bf c (right,bottom)}:  
Up/Down ratio versus log ($\Delta m^2$) for the energetic PC
events. Curves depict
predictions from $\nu_\mu \rightarrow \nu_\tau$ (dashed) and
$\nu_\mu \rightarrow \nu_{s}$ (dotted) oscillation scenarios. The
observed ratio $\pm 1 \sigma$ defines the horizontal band allowed by the data.}}
\label{fig:fig12}
\end{figure}

	More generally, the absence of coupling to the Z$^0$ for sterile 
neutrinos means that their effective potential in matter differs from that 
for $\nu_\mu$ neutrinos. The difference in respective matter potentials 
\index{matter potentials} can be written

\begin{equation}
	V_{\mu} - V_s = \mp \sqrt{2} G_F \frac{N_n}{2},
\end{equation}

\noindent
where the difference is negative (positive) for neutrinos (antineutrinos) 
and $N_n$ is the neutron number density.
For high energy $\nu_\mu$'s traversing matter, the existence of this potential
difference causes $\nu_\mu \rightarrow \nu_{s}$ 
\index{$\nu_\mu \rightarrow \nu_s$}
to be suppressed relative to 
$\nu_\mu \rightarrow \nu_\tau$ 
\index{$\nu_\mu \rightarrow \nu_\tau$} in a way that may be discernible 
\cite{Liu,Lipari3}. To elucidate the effect, the neutrinos of lower energy 
observed in Super-K \index{SuperKamiokande}
can be used to establish a baseline for neutrino 
oscillation behavior. Fig.~\ref{fig:fig12}a shows allowed regions 
\index{allowed regions} in the 
parameter plane obtained from three different oscillation fits to the FC 
single-ring events. The allowed region for $\nu_\mu \rightarrow \nu_\tau$ 
oscillations (Fig.~\ref{fig:fig12}a-top) is as described previously with 
(1.0, $3.0\times10^{-3}$ eV$^2$) as best values.
For $\nu_\mu \rightarrow \nu_s$ there are two solutions with two 
allowed regions (Fig.~\ref{fig:fig12}a-middle, bottom); 
these arise from the two possibilities with the sign of 
the mass squared difference which occurs between mass eigenstates 
involved in the mixing. However the allowed regions are found to be very 
similar in all three cases, with fits of comparable quality, consequently 
$\nu_\mu \rightarrow \nu_s$ is practically indistinguishable from 
$\nu_\mu \rightarrow \nu_\tau$ for the FC events. 

For higher energy neutrinos however this `degeneracy' can be altered by 
matter traversal, a possibility which can be seen by examining oscillation 
phenomenology appropriate for neutrinos moving through matter of uniform 
density. Interestingly, matter effects for $\nu$ oscillations can be 
formulated in a way which is look-alike to phenomenology for vacuum 
oscillations \cite{Lipari3,Rosen}. One feature is that the mixing angle for 
vacuum oscillations $\sin^2 2\theta$ goes over to $\sin^2 2\theta_m$ for 
oscillations in matter

\begin{equation}
 \sin^2 2\theta_m = \frac{\sin^2 2\theta}{\sin^2 2\theta + (D-\cos 2\theta)^2}
,
\end{equation}

\noindent where $D$ is proportional to $E_\nu$ and to the difference 
in effective potentials in matter for the mixing neutrino flavors

\begin{equation}
	D = \frac{2E_\nu V_{\alpha\beta}}{\Delta m^2} \mbox{ }
\mbox{ where }
V_{\alpha\beta} \equiv V_\alpha - V_\beta.
\label{eqn:DEquation}
\end{equation}

For $\nu_\mu \rightarrow \nu_\tau$, \index{$\nu_\mu \rightarrow \nu_\tau$}
$V_{\mu \tau} = 0$ and hence $D=0$, 
consequently matter traversal produces no effect on this oscillation. For
$\nu_\mu \rightarrow \nu_{s}$ \index{$\nu_\mu \rightarrow \nu_s$}
however, and 
for $\nu_\mu \rightarrow \nu_e$ \index{$\nu_\mu \rightarrow \nu_e$}
as well, $V_{\mu s}$ ($V_{\mu e}$) is non-zero,
and is in fact sign-dependent since anti-neutrinos are affected differently 
than are neutrinos. Then $D$ is non-zero and acquires a sizable magnitude at 
high $E_\nu$. Because of its occurrence in the denominator of
$\sin^2 2\theta_m$, it acts to suppress $\nu_\mu \rightarrow \nu_s$ at high 
energies, a suppression which is absent for $\nu_\mu \rightarrow \nu_\tau$.

To test for occurrence of matter-induced oscillation suppression, 
SuperKamiokande \index{SuperKamiokande}
has examined two different high energy neutrino samples. The 
first sample consists of PC events for which the number of photo-electrons 
from each event exceeds 45,000. This is equivalent to requiring that 
$E_{vis} > 5$ GeV; the sample thus obtained has $<E_\nu> \sim 25$ GeV. The 
zenith angle distribution for these events is shown in Fig.~\ref{fig:fig12}b.
The data (solid circles) are binned in $\cos \theta_z$ from -1.0 to 1.0. Shown 
superposed are the predictions \cite{Lipari3} from $\nu_\mu \rightarrow \nu_\tau$ and for
$\nu_\mu \rightarrow \nu_{s}$, with $\Delta m^2$ and $\sin^2 2\theta$ set to 
the values inferred from the FC events. A difference between these 
distributions is apparent for $\cos \theta_z$ below -0.2. Neutrinos which 
initiate events in this region travel thousands of kilometers through the 
Earth, and thus experience matter effects. For the $\nu_{sterile}$ oscillation
case, matter effects suppress the $\nu_\mu \rightarrow \nu_s$ oscillation, 
consequently fewer $\nu_\mu$'s ``disappear''. The expectation therefore lies 
above the curve for $\nu_\mu \rightarrow \nu_\tau$. Interestingly, it also 
lies above the data.

	To quantify the difference, an up-down ratio is used. Here,
the number of $\nu_\mu$'s which are upward-going (and consequently subject 
to matter suppression for the $\nu_\mu \rightarrow \nu_s$ case) is compared 
to the downward-going flux which, at high energies, is not affected by 
oscillations: 
$ (N_{up} (\cos \theta < -0.4) / N_{down} (\cos \theta > +0.4))_{DATA} =
0.50 \pm 0.12 \pm 0.01.$ In comparison, the ratio $N_{up}/N_{down}$ is 
is $0.94 \pm 0.04$ from the null oscillation MC. 
Assuming $\sin^2 2\theta = 1.0$, this ratio can be plotted versus 
$\Delta m^2$ as in Fig.~\ref{fig:fig12}c.
The region allowed by the data corresponds to the horizontal band centered at 
$N_{up}/N_{down} = 0.50$ with boundaries at $\pm1 \sigma$. Curves obtained from 
the $\nu_\mu \rightarrow \nu_\tau$ \index{$\nu_\mu \rightarrow \nu_\tau$}
and $\nu_\mu \rightarrow \nu_s$ 
\index{$\nu_\mu \rightarrow \nu_s$}
scenarios are
also drawn. For $\nu_\mu \rightarrow \nu_\tau$, the predicted curve falls
within the band allowed by the data for plausible values of $\Delta m^2$ which
include the best fit value obtained with the FC events. For 
$\nu_\mu \rightarrow \nu_s$ however, the scenario curve lies above the allowed
band, only ``entering'' at a $\Delta m^2$ which is higher than the FC best fit value.

\begin{figure}[htb]
\centerline{\epsfig{file=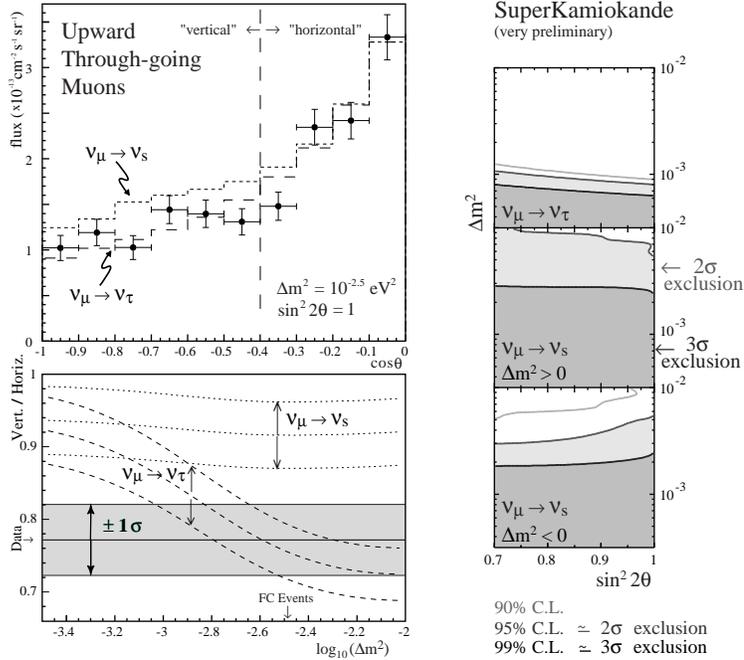,height=3.5in}}
\caption{{\footnotesize {\bf a (left,top)}: Flux of upward through-going muons
versus $\cos \theta_z$ in Super-K. 
The histograms depict $\nu_\mu \rightarrow \nu_\tau$ and
$\nu_\mu \rightarrow \nu_s$ solutions. 
{\bf b (left,bottom)}: Vertical/Horizontal ratio versus $\log (\Delta m^2)$ 
for muons of 13{\bf a}, with predictions from 
$\nu_\mu \rightarrow \nu_\tau$ (dashed band) and
$\nu_\mu \rightarrow \nu_s$ (dotted band) oscillation scenarios.
{\bf c (right)}:  
Exclusion regions for $\nu_\mu \rightarrow \nu_\tau$ and
$\nu_\mu \rightarrow \nu_s$ scenarios obtained by 
fitting to Up/Down and Vertical/Horizontal ratios from
energetic PC and through-going muon samples, to be compared to the allowed 
regions of Fig.~\ref{fig:fig12}a.}}
\label{fig:fig13}
\end{figure}

	A similar pattern is found with zenith angles for upward through-going
muons in Super-K, \index{SuperKamiokande}
a sample for which $<E_\nu>$ $\simeq~100$ GeV. 
Figure~\ref{fig:fig13}a shows the $\cos \theta_z$ distribution of that sample,
with $\nu_\mu \rightarrow \nu_\tau$ and for $\nu_\mu \rightarrow \nu_s$
shown superposed. As observed with energetic PCs, matter suppression for 
$\nu_\mu \rightarrow \nu_s$ places the prediction (at $\sin^2 2\theta$, 
$\Delta m^2$ from the FC events) above the data for angles of incidence 
corresponding to large path lengths through the Earth. To quantify the 
differences in scenarios here, the data distribution is separated into 
``horizontal'' ($\cos \theta > -0.4$) and ``vertical'' ($\cos \theta < -0.4$) 
events and the vertical to horizontal ratio is calculated:
$(N_{vertical}/N_{horizontal})_{DATA} = 0.77 \pm 0.05 \pm 0.01.$
As shown in Fig.~\ref{fig:fig13}b, there is no value of $\Delta m^2$ 
for which the $\nu_\mu \rightarrow \nu_s$ scenarios predict $N_{v}/N_{h}$ 
in the value range indicated by the through-going muon data.

The difference between data and MC predictions for the three 
oscillations scenarios can be evaluated using a chi-square function
$\chi^2 = \chi^{2}_{PC} + \chi^{2}_{thru-\mu}$ where $\chi^{2}_{PC}$ contains 
the difference for the up/down ratio and $\chi^{2}_{thru-\mu}$ the difference 
for the vertical/horizontal ratio. Parameter regions excluded at 90\% and 99\%
CL are then deduced from $\chi^2$, corresponding approximately to 2$\sigma$ 
and 3$\sigma$ exclusion. The exclusion regions for each of the three 
oscillation scenarios are displayed as shaded areas in Fig.~\ref{fig:fig13}c. 
Comparison of these excluded regions with the oscillation-allowed regions of 
Fig.~\ref{fig:fig12}a, shows large portions of 
$\nu_\mu \rightarrow \nu_s$, \index{$\nu_\mu \rightarrow \nu_s$}
for both $\Delta m^2 >$ 0 and $<$ 0, 
to be excluded at the 2$\sigma$ level. While the observations do not as 
yet rule out $\nu_\mu \rightarrow \nu_s$, it is clear that this new approach 
by Super-K \index{SuperKamiokande}
can be steadily strengthened with more exposure.

\begin{figure}[htb]
\centerline{\epsfig{file=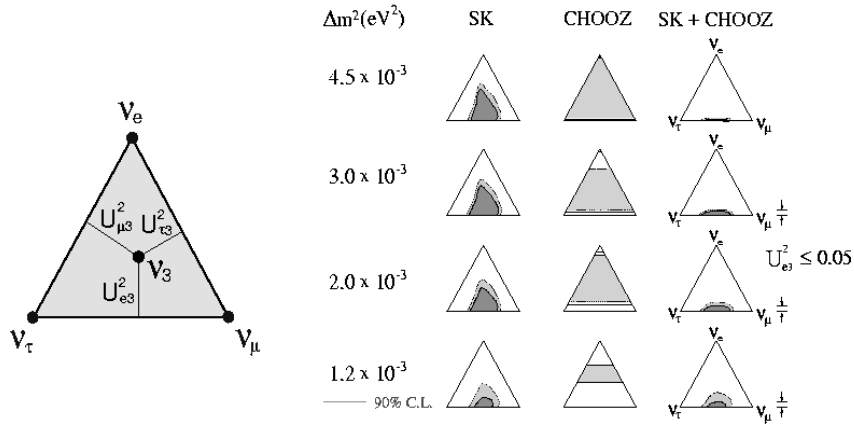,width=4.4in}}
\caption{{\footnotesize {\bf a (left)}: Triangle graph for displaying 
possible flavor compositions of the mass eigenstate $\nu_3$. 
{\bf b (right)}
Regions of allowed ($U^2_{e3}$, $U^2_{\mu 3}$, $U^2_{\tau 3}$)
versus $\Delta m^2$, obtained by fitting the Super-K and CHOOZ data via
three-flavor mixing in the approximation of one mass scale dominance \protect\cite{Fogli2}.}}
\label{fig:fig14}
\end{figure}

\section{Subdominant $\nu_\mu \leftrightarrow \nu_e$ in Three-Flavor 
Mixing}

	For a view of possibilities with subdominant oscillations, we turn 
to investigations of three-flavor mixing. A number of approaches have been 
discussed in the literature \cite{Fogli2}. Here we review a scenario for 
subdominant $\nu_\mu \rightarrow \nu_e$ 
\index{$\nu_\mu \rightarrow \nu_e$} which emerges directly from the 
approximation of one mass scale dominance. 
\index{one mass scale dominance} In this approximation it is assumed
that one of the mass eigenstates - $\nu_3$ let us say - is more massive than 
the other two, and that the lighter $\nu_{1,2}$ eigenstates are nearly 
mass-degenerate. As a result there are two mass-squares differences 
$\Delta m^2 = |m_3^2 - m_{1,2}^2|$ and $\delta m^2 = |m_2^2 - m_1^2|$ having
different magnitudes. The larger $\Delta m^2$ can be identified with the 
dominant two-state mixing observed with atmospheric oscillations, whereas 
the much smaller $\delta m^2$ relates to oscillations not observable with 
atmospheric $\nu$'s but presumably relevant to solar $\nu$'s. Up to terms of 
order $(\delta m^2)/(\Delta m^2)$ the parameter space for atmospheric 
neutrinos is spanned by $(\Delta m^2, U_{e3}^2, U_{\mu3}^2, U_{\tau3}^2)$. 
The amplitudes satisfy the unitary constraint 
$U_{e3}^2 + U_{\mu3}^2 + U_{\tau3}^2 = 1$.

For vacuum oscillations, it follows that the oscillation probability for 
transitions between flavors $\alpha$ and $\beta$ is

\begin{equation}
	P^{vac} (\nu_\alpha \leftrightarrow \nu_\beta) = 
	4 \mbox{ } U^2_{\alpha3} \cdot U^2_{\beta3} \cdot
        \sin^2 \left[ \frac{1.27 \mbox{ } \Delta m^{2} \cdot L}{E_\nu} \right].
\label{eqn:pvac}
\end{equation}

\noindent
As suggested by the amplitudes in this expression, the flavor composition of 
the massive $\nu_3$ eigenstate is the central issue:

\begin{equation}
	\nu_3 = U_{e3} \cdot \nu_e + U_{\mu 3} \cdot \nu_\mu + U_{\tau 3} 
                \cdot \nu_\tau.
\end{equation}

\noindent
For fixed $\Delta m^2$, the $\nu_3$ composition is conveniently depicted 
using the equilateral triangle construction shown in 
Fig.~\ref{fig:fig14}a for which the unitarity constraint is automatically 
incorporated.

	Fogli, Lisi, Marrone, and Scioscia have compared predictions for 
specific choices of $(\Delta m^2, U_{e3}^2, U_{\mu3}^2, U_{\tau3}^2)$ with 
Super-K \index{SuperKamiokande}
zenith angle distributions; the constraint on 
$\nu_e \rightarrow \nu_\mu$ from the CHOOZ \index{CHOOZ}
limit has also been included. They 
find that two-flavor oscillations with maximal 
$\nu_\mu \leftrightarrow \nu_\tau$ mixing works rather well \cite{Fogli2}:

\begin{equation}
\sin^2 2\theta_{\mu\tau} = 4 \mbox{ } U_{\mu3}^2 \cdot U_{\tau3}^2 \simeq 1.0
\mbox{ } \mbox{ with } \mbox{ } U_{\mu3}^2 \simeq U_{\tau3}^2 \simeq 1/2 
\mbox{ } \mbox{ and } \mbox{ } U_{e3}^2 \simeq 0. 
\end{equation}

\noindent
A small admixture of $U_{e3}^2$ is however allowed by the fits to data 
as is shown graphically in Fig.~\ref{fig:fig14}b. Here, for a relevant 
selection of $\Delta m^2$ values, the domain of $U_{f3}^2$ values allowed 
by Super-K data at 90 and 99\% CL comprise the shaded areas in the triangle 
graphs of the left-most column. Elimination of regions excluded by CHOOZ 
(see center-column triangle graphs) leaves the diminished but still 
existent allowed regions shown in the right-most column of 
Fig.~\ref{fig:fig14}b. From the height of the various allowed regions, 
it is concluded that $U_{e3}^2 \stackrel{<}{\sim} 0.05$. The expression for 
the $(\nu_\mu \rightarrow \nu_e)$ \index{$\nu_\mu \rightarrow \nu_e$}
vacuum oscillation probability follows 
immediately from Eq.~(\ref{eqn:pvac}) with $\alpha$, $\beta$ assigned to $\mu$, 
$e$ respectively. We infer from this formula that 
$P^{vac}(\nu_\mu \rightarrow \nu_e)$ can be as large as 0.10. 

\section{$P(\nu_\mu \rightarrow \nu_e)$ Amplification via Matter Resonances}
 
A $\nu_\mu \leftrightarrow \nu_e$ \index{$\nu_\mu \rightarrow \nu_e$}
oscillation of strength as indicated 
above will be hard to discern within the atmospheric flux, however we may 
get some help, as a result of amplification by matter resonances in the Earth.
Two kinds of resonance effects are possible.

\begin{figure}[htb]
\centerline{\epsfig{file=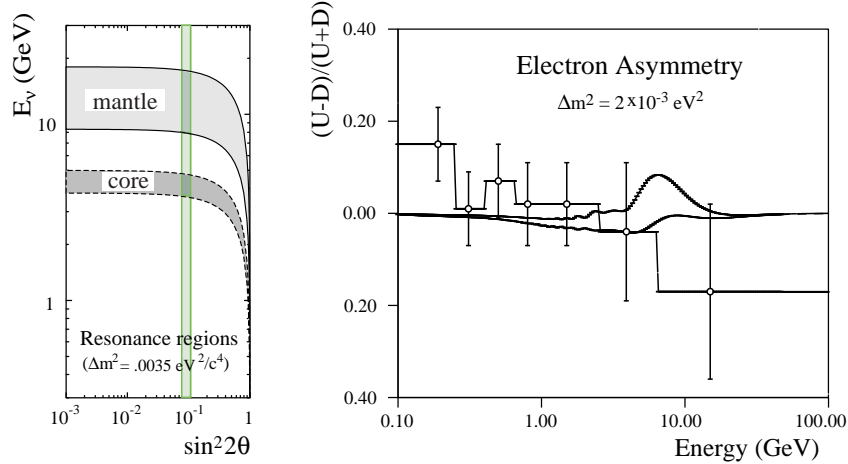,width=4.4in}}
\caption{{\footnotesize {\bf a (left)}: Values of $\sin^2 2\theta$ and $E_\nu$ for which $\nu_\mu$ to
$\nu_e$ mixing of atmospheric neutrinos could be enhanced by MSW resonance
in the Earth's mantle or core \protect\cite{Barger}. 
{\bf b (right)}: 
Possible ``bump'' in up/down asymmetry of $\nu_e$ flavor neutrinos
which could arise with MSW resonant enhancement in the Earth's mantle \protect\cite{Pantaleone}.}}
\label{fig:fig15}
\end{figure}

\noindent
One arises with the MSW resonance \index{MSW resonance}
wherein the $D$ of Eq.~(\ref{eqn:DEquation})
containing $V_{e\mu}$ approximates $\cos^2 2\theta$ with the consequence that
$\sin^2 2\theta_{m} \simeq 1.0$. Depending upon the particular values of the
mixing parameters, MSW enhancement can take place either in the terrestrial
mantle or core for the $E_{\nu}$ intervals depicted in Fig.~\ref{fig:fig15}a 
\cite{Barger}. An MSW resonance could result in a bump in the upward-going
$\nu_e$ flux at the resonance energy. This possibility has been examined 
by J. Pantaleone who proposes that the $\nu_e$ up-down asymmetry could be a 
useful discriminant, with possible outcome as illustrated in
Fig.~\ref{fig:fig15}b \cite{Pantaleone}. Of interest to 
long baseline experiments, e.g. K2K \index{K2K}
and MINOS\index{MINOS}, is the observation that MSW 
enhancement can also take place in the Earth's crust \cite{Lipari2}. 

\begin{figure}[htb]
\centerline{\epsfig{file=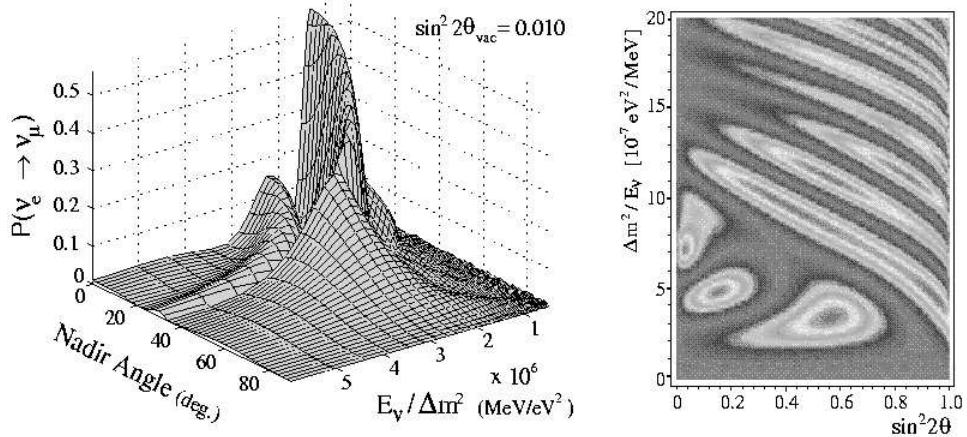,width=5.0in}}
\caption{{\footnotesize {\bf a (left)}: The probability P($\nu_{e(\mu)} \rightarrow \nu_{\mu;\tau(e)}$) 
for atmospheric neutrinos with mantle-core-mantle trajectories through the 
Earth. Such neutrinos may undergo a resonance-like enhancement (the large 
peak) which is different from an MSW enhancement. 
{\bf b (right)}: 
Regions of resonance-like enhancement for Earth core-crossing
neutrinos \protect\cite{Petcov2}.}}
\label{fig:fig16}
\end{figure}

A second and different resonance-like enhancement can take place for 
atmospheric neutrinos which cross the Earth's core. \index{Earth core}
For such neutrinos, 
having paths that cross the mantle, the core, and again the mantle, a 
complex constructive interference among the oscillation amplitudes arising 
in regions of different density is possible. The algebraic delineation of 
this effect has been presented by M.V. Chizhov and S.T. Petcov 
\cite{Petcov2}. Their work has yielded striking depictions of the transition 
probabilities as shown in Fig.~\ref{fig:fig16}. (Alternative formulations 
and interpretations have been presented; see Refs. \cite{Akhmedov2}.)

	Fig.~\ref{fig:fig16}a shows $\nu_e \leftrightarrow \nu_\mu$ 
oscillation probability as a function of nadir angle
(neutrinos at vertically upward incidence being at $0^0$) 
and of $E_{\nu}/\Delta m^2$ in units of (MeV/eV$^2$). Here the smaller 
peak at $E_{\nu}/\Delta m^2 = 2.5\times10^6$ is the MSW resonance in the 
Earth's mantle. \index{Earth mantle}
The distinctly larger structure arises with the 
mantle-core-mantle trajectories. The structures shown in
Fig.~\ref{fig:fig16}a are to be found on the ``probability island'' at 
lowest $\sin^2 2\theta$ in Fig.~\ref{fig:fig16}b. Other 
resonance structures exist at higher mixing angle values as shown.

	In the on-going underground experiments the effective 
integrations over $E_\nu$, $\cos \theta_z$, and detector resolution effects, 
which necessarily occur with the accumulation of data events, likely 
assure that matter resonance effects \index{matter resonance effects}
will be difficult to observe. Be that 
as it may, these are intriguing phenomena, underwritten by phenomenology which 
is rich and well-grounded. Their elucidation poses an interesting challenge 
for a neutrino factory to be built at a muon collider \cite{Barger,Campanelli}.

\section{Atmospheric Fluxes; Concluding Remarks}

Invaluable to all oscillation analyses with atmospheric neutrinos are 
developments which yield improved knowledge in rates and shapes of 
atmospheric flux spectra. There have been several such developments of 
recent, which we briefly remark upon here. Firstly there is the 
observation by Super-K \index{SuperKamiokande}
of the east-west anisotropy \index{east-west anisotropy}
in horizontal neutrino 
fluxes at the Kamioka site \cite{Futagami}. The $\nu_\mu$ and $\nu_e$ 
fluxes from the east are found to be depleted, as expected due to 
geomagnetic cutoff \index{geomagnetic cutoff}
of charged cosmic ray primaries and as calculated in the 
one-dimensional models of the atmospheric cascade \cite{Lipari6}. 
Just arrived ``on the scene'', are three-dimensional 
atmospheric flux calculations \index{3-D atmospheric flux}
which have been prepared independently by two
groups \cite{Battistoni}. Their arrival is timely indeed, since 3-D 
calculations are the natural framework in which to utilize the abundant 
data becoming available from balloon-borne spectrometer 
\index{balloon-borne spectrometer} experiments.
These experiments measure the primary cosmic ray flux 
\index{cosmic ray flux} and 
sample the secondary muon fluxes at a variety of depths in the atmosphere
\cite{Boezio}.

	Concerning flux-related measurements, there are ``swords-in-the-stone'' 
aplenty to tantalize the brave-of-heart. For example, no experiment to date 
has separated and compared the atmospheric anti-neutrino fluxes to the 
neutrino fluxes. It is within the capability of the underground 
experiments to distinguish $\bar{\nu}_\mu$ from $\nu_\mu$ interactions 
\cite{LoSecco}, and the $\nu$, $\bar{\nu}$ reaction cross sections are 
known. Nevertheless, the assertion by all of the flux calculations that the
ratio of $\Phi(\bar{\nu})$ to $\Phi(\nu)$ is very nearly 1:1 for either 
atmospheric neutrino flavor, remains untested. To this end, a 
$\bar{\nu}_\mu$/$\nu_\mu$ \index{$\bar{\nu}_\mu$/$\nu_\mu$}
ratio of ratios measurement would be interesting
\cite{Vannucci}. As a second example, we note the absence of measurements 
which examine variations in the atmospheric neutrino fluxes predicted to 
occur as a function of the solar cycle. \index{solar cycle}
The variations are substantial in 
the sub-GeV portion of $E_\nu$ spectra and should be more pronounced at 
northern geomagnetic latitudes \cite{Lipari4}. Such a measurement of course 
places a premium on continuous exposures which extend to a decade 
or longer; however IMB, \index{IMB} Kamiokande, \index{Kamiokande}
and Soudan 2 \index{Soudan 2} have shown 
solar-cycle-duration exposures to be attainable.

To conclude: A bonanza in neutrino oscillations research is in progress, 
driven to fever pitch by recent experimental observations with atmospheric 
neutrinos. Quite possibly, the finest nuggets are still in the ground. 
Fortunately the atmospheric beam is always on and beamline access is free; 
however the detectors required for future progress will not materialize
cheaply \cite{Aglietta}. In any case, the aura of adventure and discovery 
which now pervades the atmospheric neutrino beamline will remain for some 
time. The Dreamers and the Restless will come to try their luck, and among 
them - appearance probability of 1.0 - will be participants from this 
Symposium.

\bigskip \bigskip \bigskip
It is a pleasure to thank the organizers of the Symposium, and especially 
Helen Quinn and John Jaros, for the opportunity to give this Talk. I am 
greatly indebted to Takaaki Kajita, Serguey Petcov, Ed Kearns, John Learned, 
Francesco Ronga, Maurizio Spurio, Tomas Kafka, Jack Schneps, Maury Goodman, 
and Sandip Pakvasa for communications and discussion relating to physics 
with atmospheric neutrinos.

\renewcommand{\baselinestretch}{0.5}

\def\Discussion{
\setlength{\parskip}{0.3cm}\setlength{\parindent}{0.0cm}
     \bigskip\bigskip      {\Large {\bf Discussion}} \bigskip}

\Discussion

{\bf B.F.L. Ward (University of Tennessee):}
How do we combine the results for $\Delta m^2$ from
SuperKamiokande \index{SuperKamiokande}
and for Soudan 2, \index{Soudan 2} for example $3.5 \times 10^{-3}$ eV$^2$
and $8 \times 10^{-3}$ eV$^2$ ?

{\bf Reply:}
I would not recommend doing that. The Soudan 2 \index{Soudan 2} and
MACRO measurements are interesting as checks, with completely
different technique and systematics, on the Super-K 
\index{SuperKamiokande} result.
But since the three
determinations are in agreement and since the Super-K measurement is the 
one with predominant statistical weight, the Super-K value is the
one to be used.

\bigskip {\bf Peter Rosen (DOE):}
The evidence for oscillations from atmospheric neutrinos
is certainly impressive and the community of non-accelerator physicists is
certainly to be congratulated. To what extent can you rule out alternative
explanations? For example, Vernon Barger, Sandip Pakvasa {\it et al}. have
shown that the data can be fitted by a neutrino decay scenario.

{\bf Reply:}
The work to which you refer \cite{dis1} makes a good
case for neutrino decay \index{neutrino decay}
being viable as an alternative to neutrino
oscillations. Also, there are other explanations, e.g. flavor changing 
neutrino interactions, \index{flavor changing neutrino interactions}
which are not ruled out \cite{dis2}.

\bigskip {\bf S. Ragazzi (University of Milano):}
What do you expect to learn from the comparison of
$\Phi(\bar{\nu})$ and $\Phi(\nu)$ ?

{\bf Reply:}
The atmospheric neutrino flux calculations predict
anti-neutrino fluxes to be nearly the same as neutrino fluxes; 
\index{$\bar{\nu}_\mu$/$\nu_\mu$} this should
be tested. Although no difference is to be expected from the viewpoint of
conventional
flavor oscillations, I note that neutrino into anti-neutrino oscillation
schemes have good lineage, originating with Pontecorvo's 
\index{Pontecorvo} proposal of 1957
\cite{dis3}.

\bigskip {\bf Jasper Kirkby (CERN):} Have you looked in your data for a 
signal of $\gamma$'s produced by {\it solar} cosmic rays?  
\index{solar cosmic rays} These are produced 
by  events lasting a few days and created by energetic coronal mass
ejections from the sun.  They produce particles with peak energies of about 
100 MeV, and occasionally up to about 1 GeV.  During the most energetic events
a large ionization---equivalent to 20-30\%\ of the total annual galactic
cosmic ray flux---is dumped into the Earth's atmosphere. These events occur 
near solar maximum, which we are entering now.  I would guess Super-K 
\index{SuperKamiokande} may be 
able to detect $\nu$'s in-time with these events.  Perhaps they could even
contribute to the distortion of the solar neutrino energy spectrum we saw in 
the previous talk in the {\it hep} $\nu$ energy region.

{\bf Chang-Kee Jung (SUNY, Stony Brook):}
SuperKamiokande \index{SuperKamiokande} has examined data for solar activity
dependence for long-term periods. But we have not done so for specific
short period dependence on solar flares.

\end{document}